\documentclass[
    ,final            
	,numberedheadings 
  ]
  {aipproc}

\layoutstyle{8x11double}

\usepackage{amssymb}
\usepackage{amsmath}
\newcommand{\dd}{\mathrm{d}}

\begin{document}

\title{Simulation and Analysis Chain for Acoustic Ultra-high Energy Neutrino Detectors in Water}


\author{M.~Neff}{
  address={Friedrich-Alexander-Universit\"{a}t Erlangen-N\"{u}rnberg, Erlangen Centre for Astroparticle Physics, Erwin-Rommel-Str. 1, 91058 Erlangen, Germany}
}
\author{G.~Anton}{}
\author{A.~Enzenh\"{o}fer}{}
\author{K.~Graf}{}
\author{J.~H\"o{\ss}l}{}
\author{U.~Katz}{}
\author{R.~Lahmann}{}
\author{C.~Sieger}{}
%

\begin{abstract}
Acoustic neutrino detection is a promising approach for large-scale ultra-high energy neutrino detectors in water. In this article, a Monte Carlo simulation chain for acoustic neutrino detection devices in water will be presented. The simulation chain covers the generation of the acoustic pulse produced by a neutrino interaction and its propagation to the sensors within the detector. Currently, ambient and transient noise models for the Mediterranean Sea and simulations of the data acquisition hardware, equivalent to the one used in ANTARES/AMADEUS, are implemented. A pre-selection scheme for neutrino-like signals based on matched filtering is employed, as it is used for on-line filtering. To simulate the whole processing chain for experimental data, signal classification and acoustic source reconstruction algorithms are integrated in an analysis chain. An overview of design and capabilities of the simulation and analysis chain will be presented and preliminary studies will be discussed.
\end{abstract}

\maketitle

\setlength{\skip\footins}{.5cm}

\section{Introduction}
\label{sec1:intro}
Acoustic neutrino detection uses the effect that ultra-high energy (UHE, $E_{\nu}>10^9$\,GeV) neutrinos can produce a detectable acoustic pulse according to the thermo-acoustic model\,\cite{Askariyan1979267}. This model describes the generation of an acoustic pulse due to the local heating of the medium by a particle shower, which is caused by a UHE neutrino interaction.
The fast energy deposition by the cascade in the medium leads to a bipolar pulse.
Due to coherent emission over the cylindrical geometry of the cascade, the wave propagates through the medium in a disk-like shape perpendicular to the main axis of the cascade. Given the expected low flux of neutrinos with energies in excess of $10^9$\,GeV, a potential acoustic neutrino telescope must have large dimensions of presumably $\gtrsim  100$\,km$^3$. The attenuation length of sound in water is of the order of 1\,km for the peak spectral density of a UHE neutrino induced signal of around 10\,kHz, allowing for a sparse instrumentation. In this article, a simulation and analysis chain for an acoustic neutrino detector in water is presented. This includes the reproduction of the acoustic pulse generation, the detector properties, and the deep-sea acoustic environment. 
To simulate and evaluate the complete processing chain for experimental data, strategies for the position reconstruction and classification of acoustic signals were integrated.
Preliminary results for an effective volume and a transient-free limit setting potential of the AMADEUS\,\cite{collaboration:2010fj} acoustic detection system will be discussed.
\section{Simulation Chain}
\label{sim}
\label{sec2:sim_chain}
\begin{figure}[tb]
\centering
\includegraphics[width=0.8\columnwidth]{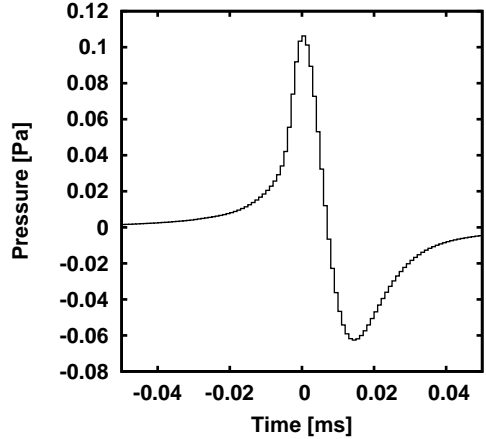}
\caption{
Simulated pressure signal for a $10^{11}$\,GeV shower as recorded at the position of the sensor. The connection line between the sensor and the shower maximum is perpendicular to the main axis of the shower; the distance between shower and sensor is 1000\,m.
\label{bip}}
\end{figure}

The simulation chain\,\cite{NeffVLVnT11} consists of the following modules, which build upon each other to create a simulated event:
\begin{itemize}
\itemsep 0pt
\item{An interaction vertex is located at a random position in a given volume around the detector and the energy and direction of the shower are set randomly within definable ranges.}
\item{The formation of the shower and the resulting acoustic signal, as generated by a UHE neutrino interaction, are simulated.}
\item{The acoustic environment of the deep sea is reproduced including both the ambient and transient noise conditions.}
\item{The data acquisition (DAQ) hardware is simulated including the system response and inherent noise of the sensors and read-out electronics.}
\item{The pre-selection scheme used on-line in the detector for data reduction is applied.}
\end{itemize}
The Monte Carlo (MC) shower is produced from a parametrisation, which is based on work by the ACoRNE collaboration\,\cite{S.Bevan:2007wd,Bevan:2009rr}. This parametrisation, which is valid up to a total shower energy of $10^{12}$\,GeV, describes the distribution of the deposited energy in the surrounding water. After the cascade has been simulated, the acoustic pulse and its propagation to the sensors within the detector are calculated. The deposited energy of the shower produces a local heating of the medium. With respect to hydrodynamical time scales, the energy deposition at time $t_0$ is instantaneous. The energy deposition $\epsilon (\mathbf{r},t)$ can be factorized into a spatial and temporal part, where the latter is approximated by using a Heaviside function. Assuming a total energy deposition $E$ in a cylindrical volume around the shower, the spatial part $\tilde{\epsilon} (\mathbf{r})$ can be expressed for the longitudinal and radial positions $z$ and $r$ in the shower as:
\begin{equation}
\tilde{\epsilon} (\mathbf{r}) = \frac{1}{E} \frac{1}{2\pi r} \frac{\dd^2E}{\dd r\dd z}
\label{eq:epsilon_part}
\end{equation}
An expression for the pressure $p$ can be derived, following\,\cite{PhysRevD.19.3293}:
\begin{equation}
p(\mathbf{r},t) = \frac{E\alpha}{4\pi C_p}\int \frac{\dd^3 r'}{R} \tilde{\epsilon}(\mathbf{r'}) \frac{\dd}{\dd t}\delta(t-\frac{R}{c_s}),
\label{eq:pressure_int_mod}
\end{equation}
where $R=|\mathbf{r} - \mathbf{r'}|$ can be approximated by the distance between the shower maximum and the sensor, $\alpha$ is the thermal expansion coefficient, $C_p$ is the specific heat capacity at constant pressure and $c_s$ the speed of sound in water. From the parametrisation of the spatial energy density distribution, which is deposited by the cascade, the pressure at a sensor at a distance $R$ from the shower can be calculated numerically\,\cite{NeffVLVnT11,Bevan:2009rr}. The pressure signal is then convoluted with the frequency dependent sound attenuation in sea water. This attenuation is based on a model by Ainslie and McColm\,\cite{1998ASAJ..103.1671A}, extended to a complex representation. 
This complete procedure calculates a UHE neutrino induced bipolar pulse at the sensor including the disk-like propagation pattern.
In Fig.~\ref{bip}, an example of a simulated pressure pulse is shown.\\
The background for acoustic neutrino detection in the deep sea consists of two different types of noise: transient and ambient noise. Transient noise signals have short duration and an amplitude that exceeds the ambient noise level. These signals can mimic bipolar pulses from neutrino interactions. Sources of transient signals can be marine mammals and anthropogenic sources such as shipping traffic. The ambient noise is mainly caused by agitation of the sea surface\,\cite{urick1986ambient}, i.e.~by wind, breaking waves, spray, and cavitations. Thus it is correlated to the weather conditions, mainly to the wind speed\,\cite{Lahmann2009255}. The model used for the simulation of the ambient noise is based on the so-called Knudsen spectra\,\cite{knudsen1948underwater}, which are adapted to the deep sea by applying attenuation effects. The wind speed distribution included in the simulation was derived from measurements at several weather stations near the coast of Marseilles, France. As reproduced by the simulation, the mean noise level $\langle\sigma_{\mathrm{noise}}\rangle$ is about 25\,mPa for the frequency range from $1\--100$\,kHz and for 95\,\% of time the noise level is smaller than $2\langle\sigma_{\mathrm{noise}}\rangle$\,\cite{Lahmann2012S216}.\\
The simulation of the DAQ hardware comprises two parts: the simulation of the sensors and of the read-out electronics. The design of the DAQ hardware is inspired by the AMADEUS\,\cite{collaboration:2010fj} project. This includes acoustic sensors using the piezo-electric effect (hydrophones) and read-out electronics to amplify and digitise the signal. The inherent noise and the system transfer function for both the sensors and the electronics have been measured in the laboratory. Sensors normally show a directional dependency of their sensitivity, therefor signal and ambient noise have to be treated separately. In this case, the incident direction of the noise is the sea surface above the detector. Signal and ambient noise are then superimposed. The inherent noise of the sensor is added. 
The resulting waveform is convoluted with the system transfer function of the read-out electronics and the corresponding inherent noise is added.\\
The output is directed to a simulation of an on-line filter system\,\cite{Neff2009S185}, which, for real data, is used to reduce the amount of data to store and to pre-select waveform samples for further off-line analysis. The filter is based on a matched filtering technique using a pre-defined bipolar pulse as reference to select signals with bipolar shape. In addition, a coincidence test between the sensors of a cluster can be performed.
\section{Analysis Chain}
\label{sec3:apps}
\begin{figure}[tb]
\centering
\includegraphics[width=0.75\columnwidth]{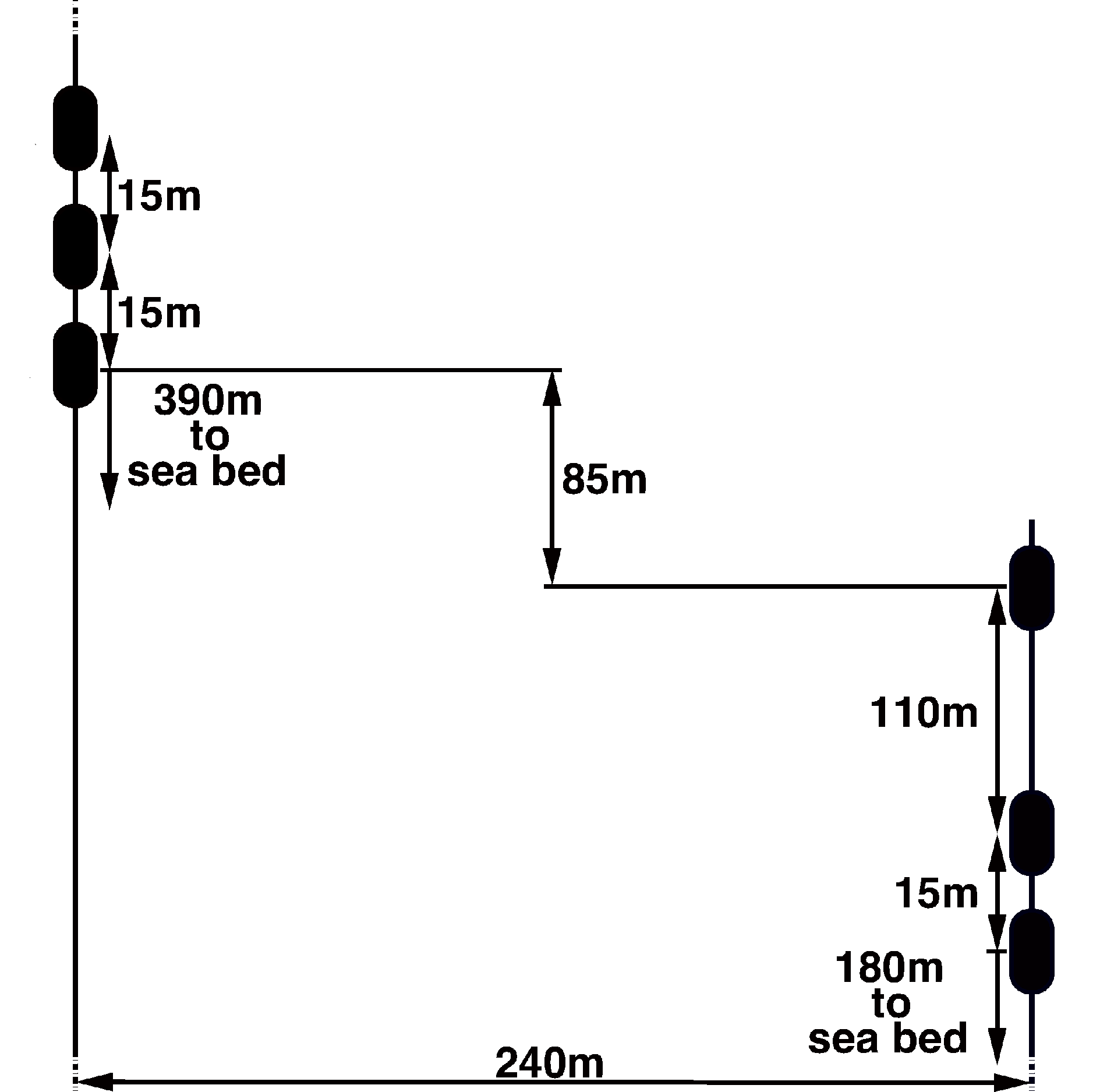}
\caption{
Sketch of the geometry of the AMADEUS detector. Sensor clusters are indicated by oval shapes and are arranged along two vertical structures. Adapted from\,\cite{collaboration:2010fj}.
\label{sketch}}
\end{figure}

\begin{figure}[tb]
\centering
\includegraphics[width=0.95\columnwidth]{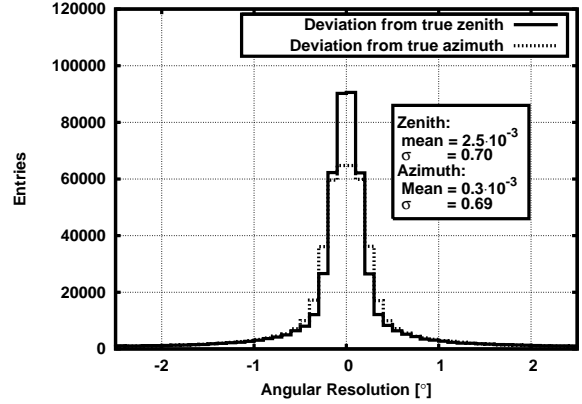}
\caption{
Angular resolution of the direction reconstruction shown for the zenith and azimuth angle. The mean of the distribution is around zero and the sigma is about 0.7$\,^{\circ}$ for both zenith and azimuth angle. The acoustic sources were generated in a cube of 5\,$\times$\,5\,$\times$\,2.5\,km$^3$ around the detector centre.
\label{direction}}
\end{figure}

\begin{figure}[tb]
\centering
\includegraphics[width=0.95\columnwidth]{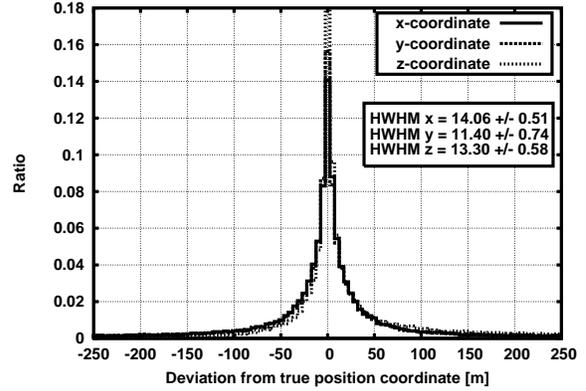}
\caption{
The distributions of the deviation of the $x$, $y$, and $z$-coordinate of the reconstructed position from the true position of the vertex. The acoustic sources were generated in a cube of 5\,$\times$\,5\,$\times$\,2.5\,km$^3$ around the detector centre.
\label{position}}
\end{figure}

The simulation chain is used to study signal classification\,\cite{Neff:2011pd} and position reconstruction algorithms\,\cite{Richardt:2009fr}. For the presented studies, a detector geometry identical to the configuration of the AMADEUS detector\,\cite{collaboration:2010fj} is used, assuming six clusters of six sensors each at fixed positions and with fixed orientation. Different sensor spacings are implemented ranging from 1\,m within clusters up to 350\,m between clusters (cf.~Fig.\,\ref{sketch}). The precise reconstruction of the arrival time of the signal is crucial for the direction and position reconstruction of the acoustic source. The arrival time is determined by cross-correlation with a pre-defined bipolar pulse. This procedure achieves a precision of about $1\,{\mathrm{\mu}} $s. The direction reconstruction is based on a least square fit of the measured arrival times at a given sensor cluster:
\begin{equation}
\min ( \sum_i (t_{\mathrm{measured}_{i}} - t_{\mathrm{expected}_{i}}(\theta,\phi))^2 ),
\end{equation}
where $i \in 1..N$ is the $i$-th sensor of a cluster of $N$ sensors, $t$ is the arrival time, and $\theta$ and $\phi$ are the zenith and azimuth angle, respectively. The acoustic sources for this analysis were generated in a cube of 5\,$\times$\,5\,$\times$\,2.5\,km$^3$ around the detector centre. The angular resolution obtained with the direction reconstruction algorithm is centred around zero with a sigma of about 0.7$\,^{\circ}$ for both zenith and azimuth angle as shown in Fig.~\ref{direction}.
\\
The position of the acoustic source is reconstructed using a ray tracing technique. If the directions were reconstructed for at least two of the sensor clusters, the intersection point (or its best approximation) of the rays starting from the sensor clusters and pointing into the reconstructed direction is searched for. In Fig.\,\ref{position}, the distributions of the deviation of the $x$, $y$, and $z$-coordinate of the reconstructed position from the true position of the vertex, as used for the simulation, is shown. The half-width at half-maximum~(HWHM) of the distributions is better than 15\,m for each coordinate.\\
The classification strategy\,\cite{Neff:2011pd} stems from machine learning algorithms trained and tested with data from the simulation. Random Forest and Boosted Trees algorithms have achieved the best results for individual sensors and clusters of sensors. For individual sensors, the classification error is of the order of 10\,\% for a well trained model. The combined results of the individual sensors in a cluster are used as new input for training. This method obtains a classification error below 2\,\%.
\section{Effective volume}
The simulation and analysis chain described above was used to simulate the data required for this study. An effective volume $V_{\mathrm{eff}}$ for the {AMADEUS} detector can be defined as:
\begin{equation}
V_{\mathrm{eff}}(E_{\nu}) = \frac{\sum_{N_{\mathrm{gen}}} \delta_{\mathrm{sel}} p(E_{\nu},\mathbf{r},\mathbf{e_{p}})}{N_{\mathrm{gen}}}V_{\mathrm{gen}}\text{,}
\label{chap4:sec3:eq:veff}
\end{equation}
where $N_\mathrm{gen}$ is the number of generated neutrino interactions in a volume $V_\mathrm{gen}$ and $p(E_{\nu},\mathbf{r},\mathbf{e_{p}})$ is the probability that the neutrino reaches the interaction vertex set in the simulation. $\delta_{\mathrm{sel}} \in \{0,1\}$ accounts for the fact that the probability only contributes to the effective volume
 if the pressure pulse corresponding to the neutrino interaction was selected by the on-line filter within a time window of 128\,${\mathrm{\mu}}$s around the expected arrival time. The probability that the neutrino reaches the vertex is given by:  
\begin{equation}
p(E_{\nu},\mathbf{r},\mathbf{e_{p}}) = e^{-d_{\mathrm{WE}}(\mathbf{r},\mathbf{e_{p}})/\lambda_{\mathrm{water}}(E_{\nu})}\,\text{,}
\label{chap4:sec3:eq:prob}
\end{equation}
where $\mathbf{r}$ is the position of the interaction vertex, $\mathbf{e_{p}}$ is the unit vector of the direction of the flight trajectory. The mean free path $\lambda_{\mathrm{water}}(E_{\nu})$ of the neutrino in water is anti-proportional to the neutrino's total cross section. The total cross section as a function of the energy $E_{\nu}$ was parameterised using values from\,\cite{Cooper-Sarkar:2011fk}. The distance $d_{\mathrm{WE}}$ is the water equivalent of the distance traveled through matter of varying density encountered by the neutrino along its flight path. For the determination of the density distribution along the flight path, the {PREM}\footnote {Preliminary earth reference model}\,\cite{Dziewonski1981297} was used to model the earth's density profile. In addition, it is assumed that the earth is covered by water of 2.5\,km depth and the detector is placed on the sea floor.
\begin{figure}[tb]
\centering
\includegraphics[width=0.95\columnwidth]
	{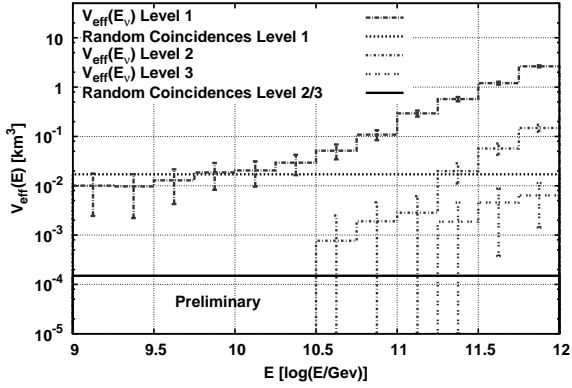}
\caption{
The effective volume of the AMADEUS detector as a function of the logarithmic neutrino energy for the three different Levels, as described in the text. Also shown are the random coincidence rates for the Levels~1 and 2/3.
\label{fig:veff}}
\end{figure}
For the calculation of the effective volume, $10^7$ neutrinos with energies uniformly distributed between $10^{9}$\,GeV and $10^{12}$\,GeV, which is the highest energy, for which the simulation is valid, were simulated. The uniform energy distribution was chosen to assure a sufficient number of events over the entire energy range. The interaction vertexes of these neutrinos were chosen in a cylindrical volume of 1200\,km$^3$ around the {AMADEUS} detector. The heading of the flight path was ranging from $0\--360\,^{\circ}$ in azimuth and from $0\--100\,^{\circ}$ in zenith\footnote{A zenith angle of $0\,^{\circ}$ corresponds to a neutrino coming from above.}. Neutrinos entering the generation volume from below the horizon will traverse an increasing amount of matter. For a zenith angle greater than $100\,^{\circ}$, the probability of a neutrino in the energy range under consideration to reach the interaction vertex is practically zero. To determine random coincidences formed by the ambient noise, a separate set of simulated data was created not containing any signals. The model used to simulate the ambient noise was described in Sec.\,\ref{sim}. The effective volume has been calculated for three different ``Levels'' describing increasingly realistic conditions and reconstruction requirements:
\begin{description}
\itemsep 0pt
\item[Level~1:]{The ambient noise is assumed to be minimal, matching sea state 0 at all times, and the coincidence requirement for the filter simulation is that at least two sensors on one storey need to respond.}
\item[Level~2:]{The complete ambient noise model and the standard on-line filter of {AMADEUS} are used, requiring at least four sensors on two storeys each.}
\item[Level~3:]{In addition to Level~2, the reconstruction of the acoustic source position is required to have a deviation from the simulated position of less than 100\,m.}
\end{description}
The results of this study are shown in Fig.\,\ref{fig:veff} for the three Levels. 
For Levels~1, 2, and 3, the effective volume is statistically incompatible with random coincidences above $1.8\cdot 10^{10}$\,GeV, $1.8\cdot 10^{11}$\,GeV, and $3.2\cdot 10^{11}$\,GeV, respectively.
The requirements of Level~1 are minimal, so this can be seen as an idealised detection threshold of the {AMADEUS} detector. The effective volume for Level~1 exceeds 2\,km$^3$ at $10^{12}$\,GeV, while for Level~2, it is about 0.1\,km$^3$ and, for Level~3, below 0.01\,km$^3$ at this energy. This result for Level~3 shows that the determination of the position of the interaction vertex has sizeable uncertainties for a small detector with an essentially two-dimensional configuration such as the {AMADEUS} system. 
\section{Transient-free limit setting potential}
\begin{figure}[tb]
\centering
\includegraphics[width=0.95\columnwidth]
	{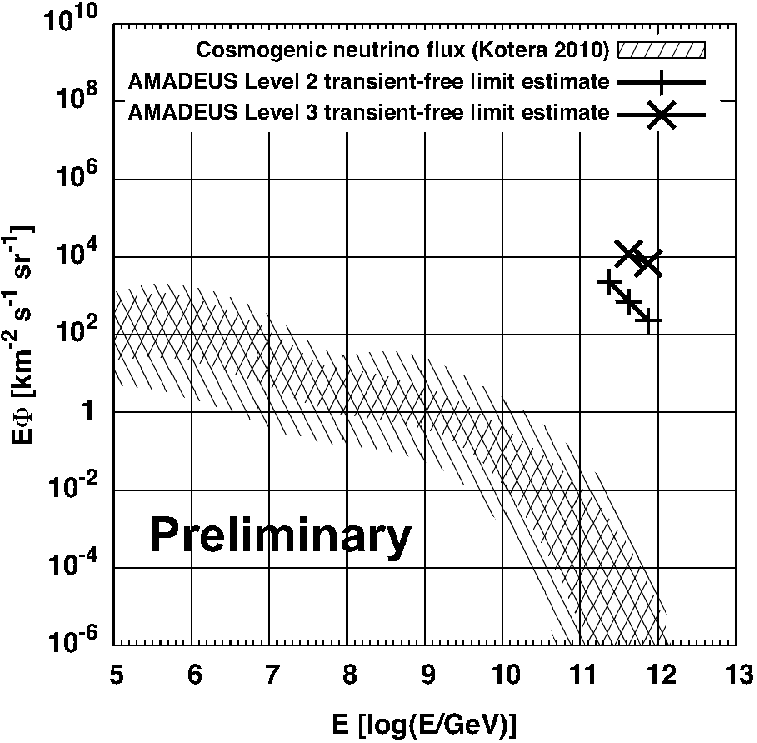}
\caption{
The transient-free limit setting potential of the {AMADEUS} detector is shown together with the theoretical neutrino flux prediction\,\cite{Kotera:2010fk} (pattern). For detail description, see the text.
\label{fig:sensitivity}}
\end{figure}

Following the approach derived in\,\cite{PhysRevD.69.013008}, a flux model independent limit estimate assuming complete transient noise suppression can be calculated as:
\begin{equation}
\Phi_{90\,\%\,\mathrm{CL}} = \frac{N_{90\,\%\,\mathrm{CL}}}{\Omega T E_{\nu} [V_{\mathrm{eff}}(E_{\nu})/\lambda(E_{\nu})]}\text{,}
\label{eq:sensitivity}
\end{equation}
where $N_{90\,\%\,\mathrm{CL}}=2.44$ for the $90\,\%$ confidence level ({CL}) assuming no background and no true signal observed\,\cite{PhysRevD.57.3873}, $\Omega$ is the solid angle, $T$ is the integrated measurement time, $V_{\mathrm{eff}}$ is the effective volume and $\lambda$ is the mean free path of the neutrino as described in the previous section. For the limit estimate of the {AMADEUS} detector to {UHE} neutrinos, the effective volume $V_{\mathrm{eff}}$ for Level~2 and Level~3 were used, as described in the previous section. The assumed integrated measurement time used for this calculation is one year and the energy range and solid angle are the same as used to calculate the effective volume. In Fig.\,\ref{fig:sensitivity}, the transient-free limit estimates of the {AMADEUS} detector are shown together with predictions for the cosmogenic neutrino flux\,\cite{Kotera:2010fk}. 
The limit estimates derived show the potential of the acoustic UHE neutrino detection technique. Clearly, the limit setting potential of the detector is limited by its small effective volume. Complete transient noise suppression is challenging and relies on a good position reconstruction to define a sensible fiducial volume. As the results for Level 3 demonstrate, the size and geometry of the AMADEUS detector are not favourable for this task.
\section{Conclusion and Outlook}
\label{sec3:c&o}
As shown, the simulation chain is capable of reproducing the aspects necessary for acoustic neutrino detection \--- from the generation of the bipolar acoustic signal to the simulation of different detector geometries and components. Also the deep-sea environment with its variable and diverse noise conditions is well replicated. Furthermore, the analysis chain for experimental data, employing position reconstruction and classification algorithms, was verified using simulations. Based on this software framework, a calculation of the effective volume and the transient-free limit setting potential of the {AMADEUS} detector was performed. 
For realistic conditions, the effective volume is around 0.1\,km$^3$ at $10^{12}$\,GeV. For a small detector such as {AMADEUS}, the calculated transient-free limit estimate is promising and encourages further studies for large-scale detectors.
%
%
\begin{theacknowledgments}
M.N. would like to thank Sean Danaher from the ACoRNE collaboration for the provided source code. This work is supported by the German government (BMBF) with grants 05A08WE1 and 05A11WE1.
\end{theacknowledgments}

%
%

\bibliographystyle{aipproc}   
%
\bibliography{biblio.bib}

\begin{thebibliography}{19}
\expandafter\ifx\csname natexlab\endcsname\relax\def\natexlab#1{#1}\fi
\providecommand{\enquote}[1]{``#1''}
\expandafter\ifx\csname url\endcsname\relax
  \def\url#1{\texttt{#1}}\fi
\expandafter\ifx\csname urlprefix\endcsname\relax\def\urlprefix{URL }\fi
\providecommand{\eprint}[2][]{\url{#2}}

\bibitem[Askariyan et~al.(1979)]{Askariyan1979267}
G.~Askariyan, et~al., \emph{Nucl. Instrum. Meth. A} \textbf{164}, 267 -- 278
  (1979).

\bibitem[Aguilar et~al.(2011)]{collaboration:2010fj}
J.~Aguilar, et~al., \emph{Nucl. Instrum. Meth. A} \textbf{626}, {128--143}
  (2011).

\bibitem[{Neff} et~al.(in publication)]{NeffVLVnT11}
M.~{Neff}, et~al., \enquote{Simulation Chain for Acoustic Ultra-high Energy
  Neutrino Detectors,} in \emph{Proceedings of the VLVnT 2011}, in publication.

\bibitem[Bevan et~al.(2007)]{S.Bevan:2007wd}
S.~Bevan, et~al., \emph{Astropart. Phys.} \textbf{28}, 366--379 (2007).

\bibitem[Bevan et~al.(2009)]{Bevan:2009rr}
S.~Bevan, et~al., \emph{Nucl. Instrum. Meth. A} \textbf{607}, 398--411 (2009).

\bibitem[Learned(1979)]{PhysRevD.19.3293}
J.~G. Learned, \emph{Phys. Rev. D} \textbf{19}, 3293--3307 (1979).

\bibitem[{Ainslie} and {McColm}(1998)]{1998ASAJ..103.1671A}
M.~A. {Ainslie}, and J.~G. {McColm}, \emph{J. Acoust. Soc. Am.} \textbf{103},
  1671--1672 (1998).

\bibitem[Urick(1986)]{urick1986ambient}
R.~Urick, \emph{Ambient noise in the sea}, Peninsula Publishing, 1986, ISBN
  9780932146137.

\bibitem[Lahmann(2009)]{Lahmann2009255}
R.~Lahmann, \enquote{Deep-sea acoustic neutrino detection and the AMADEUS
  system as a multi-purpose acoustic array,} in \emph{Proceedings of the VLVnT
  2009}, 2009, vol. 602, pp. 255 -- 261.

\bibitem[Knudsen et~al.(1948)]{knudsen1948underwater}
V.~Knudsen, R.~Alford, and J.~Emling, \emph{J. Mar. Res} \textbf{7}, 410--429
  (1948).

\bibitem[Lahmann(2012)]{Lahmann2012S216}
R.~Lahmann, \enquote{Status and recent results of the acoustic neutrino
  detection test system AMADEUS,} in \emph{Proceedings of the ARENA 2010},
  2012, vol. 662, Supplement 1, pp. S216 -- S221.

\bibitem[Neff et~al.(2009)]{Neff2009S185}
M.~Neff, et~al., \enquote{AMADEUS on-line trigger and filtering methods,} in
  \emph{Proceedings of the ARENA 2008}, 2009, vol. 604, pp. S185 -- S188.

\bibitem[Neff et~al.(2012)]{Neff:2011pd}
M.~Neff, et~al., \enquote{Signal classification for acoustic neutrino
  detection,} in \emph{Proceedings of the ARENA 2010}, 2012, vol. 662,
  Supplement 1, pp. S242 -- S245.

\bibitem[Richardt et~al.(2009)]{Richardt:2009fr}
C.~Richardt, et~al., \emph{Nucl. Instrum. Meth. Suppl. A} \textbf{604},
  189--192 (2009).

\bibitem[Cooper-Sarkar et~al.(2011)]{Cooper-Sarkar:2011fk}
A.~Cooper-Sarkar, P.~Mertsch, and S.~Sarkar, \emph{J. High Energ. Phys.}
  \textbf{2011}, 1--20 (2011).

\bibitem[Dziewonski and Anderson(1981)]{Dziewonski1981297}
A.~M. Dziewonski, and D.~L. Anderson, \emph{Phys. Earth Planet. Inter.}
  \textbf{25}, 297 -- 356 (1981).

\bibitem[Lehtinen et~al.(2004)]{PhysRevD.69.013008}
N.~G. Lehtinen, et~al., \emph{Phys. Rev. D} \textbf{69}, 013008 (2004).

\bibitem[Feldman and Cousins(1998)]{PhysRevD.57.3873}
G.~J. Feldman, and R.~D. Cousins, \emph{Phys. Rev. D} \textbf{57}, 3873--3889
  (1998).

\bibitem[Kotera et~al.(2010)]{Kotera:2010fk}
K.~Kotera, D.~Allard, and A.~Olinto, \emph{J. Cosmol. Astropart. Phys.}
  \textbf{2010}, 013 (2010).

\end{thebibliography}
%
\IfFileExists{\jobname.bbl}{}
 {\typeout{}
  \typeout{******************************************}
  \typeout{** Please run "bibtex \jobname" to optain}
  \typeout{** the bibliography and then re-run LaTeX}
  \typeout{** twice to fix the references!}
  \typeout{******************************************}
  \typeout{}
 }
\end{document}